\documentclass[
]{ceurart}

\usepackage{listings}
\usepackage{subcaption}
\usepackage{url}
\begin{document}

\copyrightyear{2026}
\copyrightclause{Copyright for this paper by its authors.
  Use permitted under Creative Commons License Attribution 4.0
  International (CC BY 4.0).}

\conference{In: R. Campos, A. Jorge, A. Jatowt, S. Bhatia, M. Litvak (eds.): Proceedings of the Text2Story'26 Workshop, Delft (The Netherlands), 29-March-2026}

\title{Semantic Interaction for Narrative Map Sensemaking: An Insight-based Evaluation}


\author[1,2]{Brian Felipe Keith-Norambuena}[%
orcid=0000-0001-5734-8962,
email=brian.keith@ucn.cl
]
\cormark[1]

\author[2]{Fausto German}[%
orcid=0009-0005-0954-4578
]

\author[3]{Eric Krokos}[%
]

\author[3]{Sarah Joseph}[%
]

\author[2]{Chris North}[%
orcid=0000-0002-8786-7103
]

\address[1]{Universidad Católica del Norte, Av. Angamos 0610, Antofagasta, 1270709, Chile}
\address[2]{Virginia Tech, 620 Drillfield Drive, Blacksburg, VA, 24061, USA}
\address[3]{U.S. Government, Washington, D.C. 20500, USA}

\cortext[1]{Corresponding author.}

\begin{abstract}
Semantic interaction (SI) enables analysts to incorporate their cognitive processes into AI models through direct manipulation of visualizations. While SI frameworks for narrative extraction have been proposed, empirical evaluations of their effectiveness remain limited. This paper presents a user study that evaluates SI for narrative map sensemaking, involving 33 participants under three conditions: a timeline baseline, a basic narrative map, and an interactive narrative map with SI capabilities. The results show that the map-based prototypes yielded more insights than the timeline baseline, with the SI-enabled condition reaching statistical significance and the basic map condition trending in the same direction. The SI-enabled condition showed the highest mean performance; differences between the map conditions were not statistically significant but showed large effect sizes ($d > 0.8$), suggesting that the study was underpowered to detect them. Qualitative analysis identified two distinct SI approaches---corrective and additive---that enable analysts to impose quality judgments and organizational structure on extracted narratives. We also find that SI users achieved comparable exploration breadth with less parameter manipulation, suggesting that SI serves as an alternative pathway for model refinement. This work provides empirical evidence that map-based representations outperform timelines for narrative sensemaking, along with qualitative insights into how analysts use SI for narrative refinement.
\end{abstract}


\maketitle

\section{Introduction}
Visual analytics (VA) seeks to support the analysis of large and complex data sets through interactive exploration~\cite{cook2005illuminating}. VA systems assist users in their sensemaking process by allowing them to interact with statistical and artificial intelligence (AI) models through visualizations to explore connections and hypotheses in the data~\cite{endert2012semantic}. Semantic interaction (SI)~\cite{endert2012semantic, bradel2014multi} enables users to incorporate their cognitive processes into the underlying computational models through intuitive interactions with visualizations. Rather than requiring direct parameter manipulation, SI allows analysts to perform natural interactions that are interpreted and translated into model updates. This approach has shown promise in various VA domains, and recent work has developed SI frameworks specifically for narrative map extraction~\cite{keith2023iui}.

Narrative maps---graph-based representations that capture connections between events in a story---are a useful tool for narrative sensemaking~\cite{keith2020maps}. However, while SI frameworks for narrative maps have been proposed, empirical evaluations of their effectiveness in supporting real sensemaking tasks remain limited. Understanding how analysts actually use SI capabilities, what strategies they develop and whether SI leads to better analytical outcomes requires user studies.

In this paper, we present a user study that evaluates semantic interaction for narrative map sensemaking. We compare an SI-enabled prototype with baseline conditions using 33 participants examining narratives from the 2021 Cuban protests. Our study employs an insight-based evaluation methodology~\cite{north2011comparison} to assess how effectively SI supports the sensemaking process. Our contributions include: \textbf{(1)} empirical evidence that map-based visualizations significantly outperform timeline baselines, with suggestive but not statistically significant additional benefits from SI; \textbf{(2)} identification of distinct SI strategies---corrective and additive---that analysts employ for narrative refinement; \textbf{(3)} analysis of SI's effect on exploration behavior, suggesting it serves as an alternative to parameter-based refinement; and \textbf{(4)} practical insights for SI system design derived from observed user strategies and perceptions.

The rest of this paper is organized as follows. We review related work in Section~\ref{sec:related} and describe the SI-enabled narrative map system in Section~\ref{sec:system}. Section~\ref{sec:study} presents the design of the user study, followed by the results in Section~\ref{sec:results}. We discuss implications and limitations in Section~\ref{sec:discussion} and conclude in Section~\ref{sec:conclusions}.

\section{Related Work}
\label{sec:related}
In this work, we combine two lines of research: \textit{narrative maps}, which computationally extract graph-based story representations from document collections, and \textit{semantic interaction}, which allows analysts to refine underlying models by directly manipulating visualizations. We review each in turn, along with the evaluation methodology we adopt.

\textbf{Narrative Extraction and Visualization.}
Narratives can be defined as systems of interrelated storylines that share coherent themes~\cite{halverson2011master}. Computational narrative extraction typically relies on event-based models using discrete structures such as timelines or directed acyclic graphs (DAGs)~\cite{keith2023survey}. Narrative maps~\cite{keith2020maps} represent narratives as DAGs where nodes are events and edges represent narrative connections, similar to other narrative representations~\cite{shahaf2010connecting, ansah2019graph}. Previous work has established the utility of narrative maps for sensemaking tasks~\cite{keith2022design,keith2023iui}. However, empirical evaluations comparing interactive narrative map systems with baselines in realistic sensemaking scenarios remain limited.

\textbf{Semantic Interaction.} SI~\cite{endert2012semantic} enables users to communicate analytical intent through intuitive visualization interactions rather than direct parameter manipulation. SI has been applied in document analysis~\cite{endert2012semantic}, spatialization~\cite{bradel2014multi}, high-dimensional data exploration~\cite{dowling2018bidirectional}, and text analytics~\cite{self2018observation, hu2013semantics}. Keith et al.~\cite{keith2023iui} proposed a mixed multi-model SI framework specifically for narrative maps, addressing challenges of combining discrete graph structures with continuous embedding spaces. Our work builds on this framework by providing empirical evaluation of its effectiveness.

\textbf{Evaluating Visual Analytics Systems.} Insight-based evaluation~\cite{saraiya2005insight, north2011comparison} assesses VA systems by measuring the insights analysts generate rather than task completion metrics. This approach better captures the effectiveness of the system for exploratory sensemaking~\cite{guo2015case} and accommodates diverse analytical strategies~\cite{alves2022personality}. We adopt this methodology to evaluate the effectiveness of SI for narrative sensemaking.

\section{System Description}
\label{sec:system}
We evaluate a narrative map system with semantic interaction capabilities, built upon the mixed multi-model SI framework proposed by Keith et al.~\cite{keith2023iui}. This section briefly describes the key components of the system that are relevant to understanding the user study.

\textbf{Narrative Extraction.} Our work falls within the event-based narrative extraction paradigm~\cite{keith2023survey}, where each document represents a single event and a narrative is a temporally ordered, thematically coherent sequence of such document-events. The system extracts narrative maps from news article collections following the approach of Keith and Mitra~\cite{keith2020maps}. Articles are represented using sentence-level Transformer embeddings (all-MiniLM-L6-v2) aggregated to produce document-level representations, projected to 2D using Uniform Manifold Approximation and Projection (UMAP)~\cite{mcinnes2018umap} for visualization, and clustered using Hierarchical Density-Based Spatial Clustering of Applications with Noise (HDBSCAN)~\cite{mcinnes2017hdbscan} to identify topical groups. A linear programming optimization then constructs the narrative map as a directed acyclic graph by maximizing coherence---a measure combining content similarity (via embedding cosine similarity) and topical relatedness (based on cluster assignments)---subject to coverage constraints that ensure that the narrative spans diverse topics. Users can adjust three parameters: \textit{map size} (expected length of the main story), \textit{coverage} (topic diversity), and \textit{temporal sensitivity} (preference for temporally proximate connections).

\textbf{Semantic Interaction Capabilities.} The SI-enabled prototype allows analysts to influence the underlying models through direct manipulation. Analysts can perform \textit{edge interactions} to add or remove connections between events, signaling which relationships are meaningful or spurious. \textit{Node interactions} allow pinning events to ensure their inclusion or removing irrelevant events from the narrative. Finally, \textit{cluster interactions} enable analysts to define custom topical clusters by grouping events, directly influencing the topical similarity component of the coherence measure. These interactions are translated into model updates through inverse transformations that modify cluster assignments and impose structural constraints~\cite{keith2023iui}. Updates can accumulate across interactions, allowing the model to incrementally learn from the analyst's reasoning.

\section{User Study}
\label{sec:study}
\textbf{Evaluation Methodology.} We evaluated our SI framework through a user study using an insight-based evaluation methodology~\cite{saraiya2005insight, north2011comparison}. In contrast to traditional benchmark evaluations, insight-based evaluations capture how analysts use VA systems and arrive at insights, providing a more realistic assessment of the effectiveness of the system for exploratory sensemaking tasks~\cite{guo2015case,alves2022personality}.

Participants were asked to identify as many insights as possible about the causes and effects of the 2021 Cuban protests and how they relate to each other. This open-ended task gives analysts freedom to explore the dataset and approximates a realistic narrative sensemaking scenario. We consider our SI system effective if users are able to gather more insights with it compared to baselines.

\textbf{Data and Participants.} We used news data covering the 2021 Cuban protests~\cite{keith2023iui}. The dataset contained 160 news articles from multiple sources across the political spectrum, providing an appropriate level of complexity for evaluating narrative sensemaking capabilities. We recruited 36 graduate and undergraduate students, primarily in Computer Science, with additional participants in Communications, National Security, and Engineering. After excluding dropouts and participants who failed attention checks, the final sample consisted of 33 participants. Most participants reported minimal prior knowledge of the 2021 Cuban protests (mean familiarity = 1.5 on a 5-point scale).

This study was approved by the Institutional Review Board at Virginia Tech (IRB-21-929). All participants provided their informed consent prior to participation through a consent information sheet. Participants were compensated with gift cards upon completion of the session. Screen and audio recordings collected during think-aloud sessions were stored securely and accessible only to the research team; all data reported in this article have been anonymized.

\textbf{Experimental Conditions.} Participants were randomly assigned to one of three conditions. The \textbf{Timeline (TL)} condition served as the baseline, providing a timeline-based prototype with search functionality that represents a traditional approach to exploring temporal document collections. The \textbf{Basic Map (BM)} condition provided a narrative map prototype without SI capabilities---users could adjust parameters (size, coverage, temporal sensitivity) but could not influence the underlying models through interactions. The \textbf{Interactive Map (IM)} condition provided a narrative map prototype with full SI capabilities, including edge manipulation, node pinning, and custom clustering. Participants received a 15-minute training session using a separate COVID-19 dataset before completing the main task with their assigned prototype. The final distribution was 10 participants in TL, 13 in BM, and 10 in IM. They were given 1 hour to explore the Cuban protests data and write a report documenting their insights.

\textbf{Measures and Insight Coding.} We collected both quantitative and qualitative data. Quantitative measures included the number of unique insights identified, self-reported familiarity before and after the task to compute effectiveness as the fraction of the knowledge gap closed, and Likert-scale survey responses on usability, SI effectiveness, and trust. We employed a think-aloud protocol, capturing screen activity and voice recordings throughout the session. Complete interaction logs recorded user interactions, parameter changes, and events read. Insights were coded from three sources: spoken insights from session transcripts, written insights from participant reports, and implicit insights inferred from interactions.

Insight coding was performed by the first author, who iteratively developed the hierarchical coding scheme (Figure~\ref{fig:coding}) during the analysis. The coded insights were then presented to domain experts among the co-authors for review; they provided feedback on whether categories were sensible and flagged questionable codings for re-examination. We did not perform formal inter-rater reliability assessment, which we acknowledge as a limitation. However, standardized practices for coding consistency in insight-based evaluations remain underdeveloped~\cite{gomez2014insight}, in part because there is no unified definition of what constitutes an insight~\cite{battle2023exactly}. Our quantitative insight counts are derived from the structured hierarchical codes, and the overall patterns were consistent across both high-level and detailed insight measures.

Regarding implicit insights---those inferred from user interactions such as removing biased sources or creating topical clusters---we note that these were coded conservatively based on what each interaction logically implied. However, because the IM condition affords richer interactions than BM or TL, implicit insights may disproportionately favor the IM condition. To assess whether this disproportionately favored the IM condition, we repeated the analysis excluding interaction-inferred insights.
The general pattern held: IM significantly outperformed TL for both high-level ($p = 0.028$) and detailed ($p = 0.022$) insights,confirming that the results are not driven by implicit insight coding. Nevertheless, we recommend that future insight-based evaluations of interactive systems explicitly separate interaction-inferred insights from spoken and written insights to enable fair cross-condition comparison.

\section{Results}
\label{sec:results}
\subsection{Insight Generation}
\begin{figure}[!htb]
    \centering
    \includegraphics[width=0.95\textwidth]{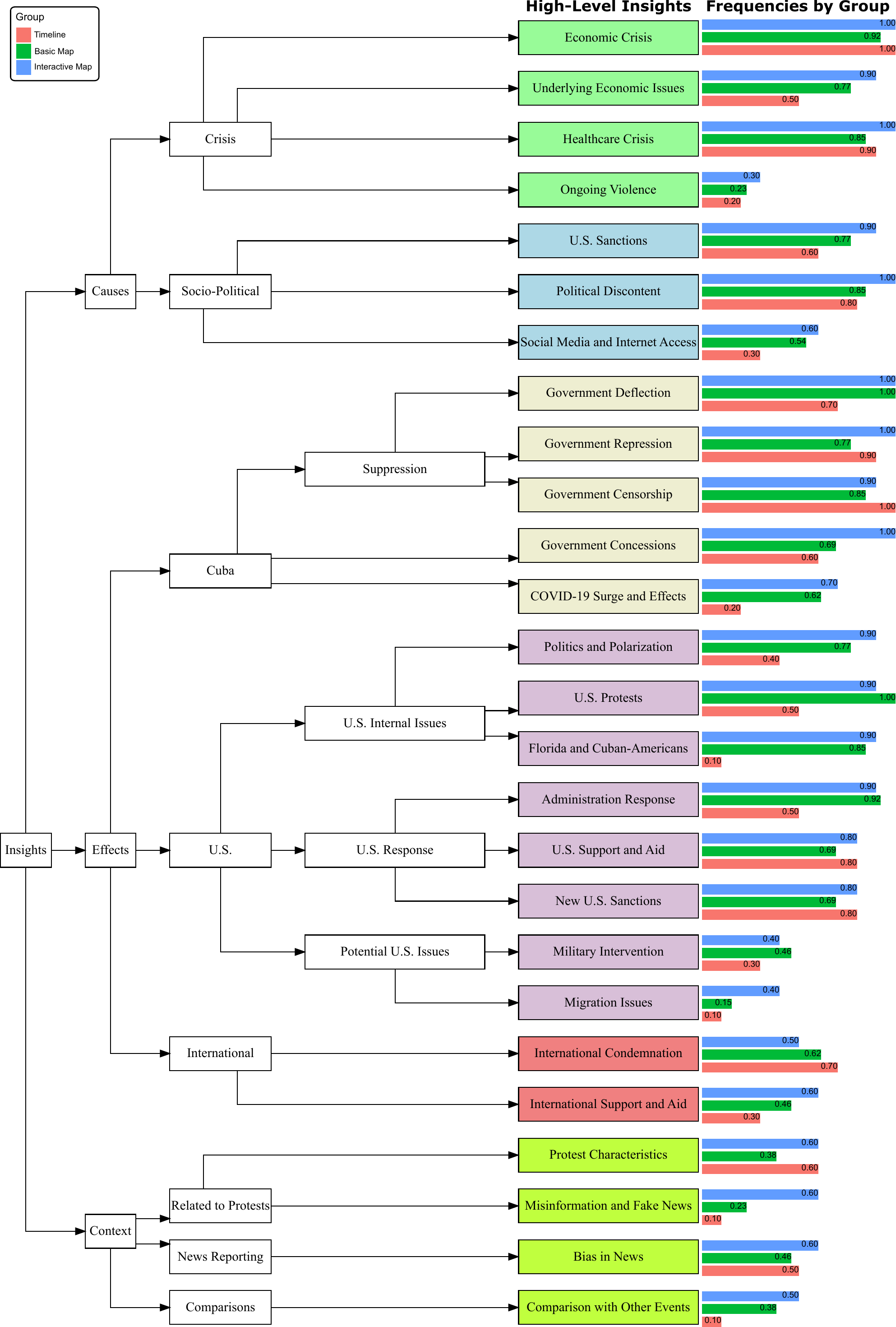}
    \caption{Hierarchical coding scheme for insights. The leaves represent high-level insights and include histograms showing the frequencies by group (TL, BM, IM). Colors distinguish causes (green), effects in Cuba (red), effects in the U.S. and internationally (blue), and contextual insights (teal).}
    \label{fig:coding}
\end{figure}

\begin{figure}[!htb]
    \centering
    \includegraphics[width=1.0\textwidth]{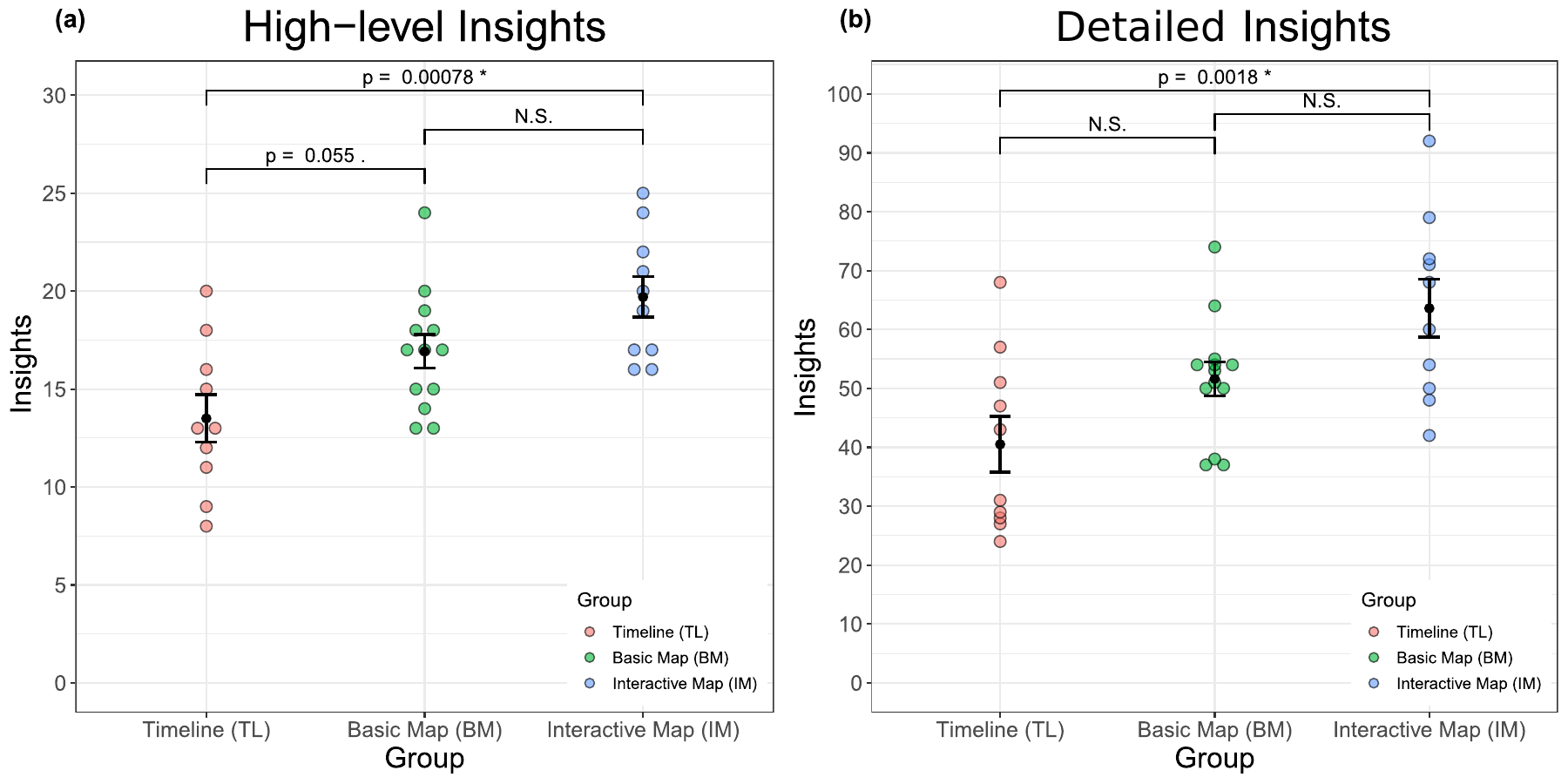}
    \caption{\textbf{(a)} High-level insights results. \textbf{(b)} Detailed insights results. (N.S.) denotes no significant difference. (.) denotes marginal significance ($p \in [0.05, 0.1)$). (*) denotes significance ($p < 0.05$). All reported $p$-values are from post-hoc pairwise comparisons following significant ANOVA results. The bars represent standard errors.}
    \label{fig:insights}
\end{figure}

Figure~\ref{fig:coding} shows the hierarchical coding scheme used to categorize insights and the frequency distributions between conditions. Participants with the IM prototype generated more insights on average compared to both the BM and TL conditions. Figure~\ref{fig:insights} shows the comparison between experimental conditions in terms of quantity of high-level and detailed insights.

For high-level insights, the average counts were: TL = 13.5, BM = 16.9, IM = 19.7. The analysis of variance (ANOVA) revealed a significant difference between groups ($p = 0.00117$). Post-hoc Tukey HSD tests showed that IM significantly outperformed TL ($p = 0.00078$), with a marginally significant difference between BM and TL ($p = 0.055$). The IM-BM difference was not significant ($p = 0.139$). Effect sizes were large for all pairwise comparisons: IM--TL $d = 1.75$, BM--TL $d = 1.01$, and IM--BM $d = 0.88$. 

For detailed insights, the pattern was similar: TL = 40.5, BM = 51.6, IM = 63.6. The ANOVA showed a significant effect ($p = 0.00267$), with IM significantly outperforming TL ($p = 0.0018$). However, neither the BM-TL comparison ($p = 0.143$) nor the IM-BM comparison ($p = 0.106$) were significant for detailed insights. Effect sizes for detailed insights followed the same pattern: IM--TL $d = 1.51$, BM--TL $d = 0.88$, and IM--BM $d = 0.93$.

These results provide evidence that narrative map representations support sensemaking compared to the timeline baseline, with the clearest benefits in the SI-enabled condition. The lack of significant differences between IM and BM means we cannot conclusively attribute the IM advantage to SI specifically. However, all pairwise effect sizes were large ($d > 0.8$), indicating that non--significant comparisons between IM and BM reflect limited statistical power rather than small effects. A post-hoc power estimate suggests that approximately 20 participants per group would be needed to detect the observed IM--BM differences at conventional significance levels ($\alpha = 0.05$, power $= 0.80$). The pattern of results is therefore consistent with additional benefits from SI that our sample was unable to confirm statistically.

\subsection{Effectiveness and Reading Behavior}
The average effectiveness (fraction of knowledge gap closed, measured via familiarity ratings) was: TL = 0.68, BM = 0.79, IM = 0.83. The ANOVA revealed a marginally significant effect ($p = 0.078$). Post-hoc Tukey HSD tests showed a marginally significant difference between IM and TL ($p = 0.072$); other pairwise comparisons did not reach significance (BM-TL: $p = 0.189$; IM-BM: $p = 0.621$). The difference in IM-TL suggests that the SI-enabled prototype may help users develop a more complete understanding of the narrative, although the intermediate BM performance indicates that map-based representation contributes to this effect.

\begin{figure}[!htb]
    \centering
    \includegraphics[width=1.0\textwidth]{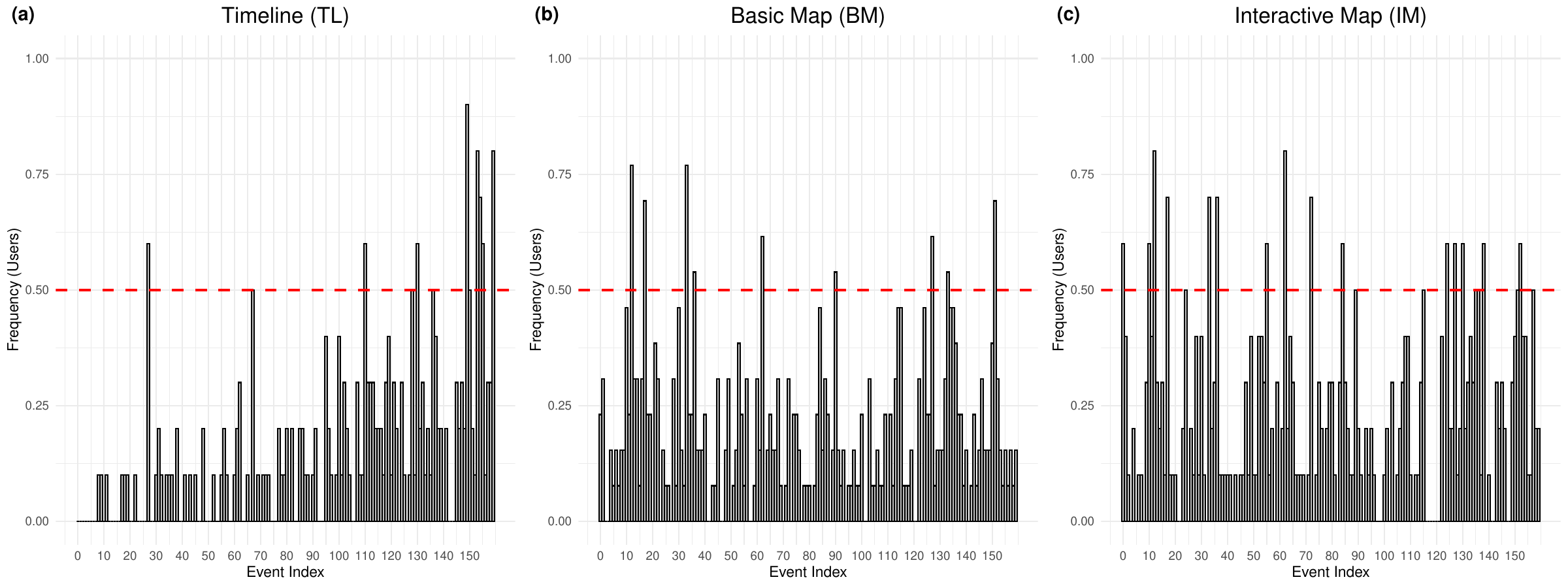}
    \caption{Reading distributions by condition. The red line indicates ideal uniform distribution.}
    \label{fig:reading}
\end{figure}

Regarding reading behavior, map-based prototypes led to a more uniform coverage of the dataset (see distributions in Figure \ref{fig:reading}). The TL condition showed a reading distribution skewed toward later events (due to search results prioritizing recent events, combined with participants focusing on early result pages), while both map conditions showed roughly uniform distributions across the timeline. Pairwise comparisons confirmed that both map conditions differed significantly from TL ($p < 0.005$) but not from each other.

Table~\ref{tab:si_survey} shows the user evaluation of SI. The participants found the SI capabilities useful ($M = 4.20$) and agreed that the system learned from their interactions ($M = 4.20$). Interactions were considered relevant ($M = 3.90$). Trust in the model was positive but somewhat lower ($M = 3.80$). All ratings were positive (above the 3.0 midpoint). The trust ratings were numerically lower than the usefulness ratings, although this difference did not reach statistical significance (paired $t(9) = 1.50$, $p = 0.168$).

\begin{table}[!htb]
\caption{Survey results for SI components (5-point Likert scale, IM condition only, $n = 10$).}
\label{tab:si_survey}
\centering
\begin{tabular}{lcccccc}
\toprule
Item & 1 & 2 & 3 & 4 & 5 & M \\
\midrule
Interaction Relevance & 0 & 0 & 1 & 9 & 0 & 3.90 \\
Learning Model & 0 & 0 & 0 & 8 & 2 & 4.20 \\
Usefulness & 0 & 0 & 1 & 6 & 3 & 4.20 \\
Model Trust & 0 & 0 & 4 & 4 & 2 & 3.80 \\
\bottomrule
\end{tabular}
\end{table}

\subsection{User Strategies}
We identified general exploration strategies employed by participants with map-based prototypes (Table~\ref{tab:general-strategies}). The two dominant approaches were \textit{Single Map}---deeply exploring one narrative representation---and \textit{Parametric Exploration}---generating multiple maps with varied parameters. Participants often shifted between these strategies as they learned the system, for example starting with single map focus before moving to parametric exploration. These approaches reflect different orientations toward narrative sensemaking: depth-first exploration of a fixed narrative structure versus breadth-first comparison across alternative representations.

\begin{table}[!htb]
\caption{General exploration strategies used by participants with map-based prototypes (BM and IM conditions).}
\label{tab:general-strategies}
\centering
\small
\begin{tabular}{p{4cm}p{11cm}}
\toprule
\textbf{Strategy} & \textbf{Description} \\
\midrule
Single Map & Focused on a single map, extracting as much information as possible from one representation. \\
Parametric Exploration & Generated several maps with different parameters or starting events to extract information broadly. \\
Comprehensive Timeline & Subtype of Single Map; generated a large timeline-like map with a long main storyline and few side stories. \\
Limited Parametric Exploration & Subtype of Parametric Exploration; generated only a few maps, intermediate between the two main strategies. \\
Parametric Exploration with Corrections & Attempted to correct the map during parametric exploration, even in BM where SI was unavailable. \\
Incremental Maps & Iteratively increased map size, starting from a small overview and building toward complexity. \\
Sliding Window & Iteratively changed the starting event in a sliding window manner (e.g., event 0, then 10, then 20). \\
Increasing Coverage & Iteratively increased map coverage, starting with low coverage and expanding to cover more topics. \\
Parametric Exploration with Keyword Search & Complemented parametric exploration with keyword search to explore unselected events. \\
Headline Exploration & Focused solely on headlines without reading events until the big picture was established. \\
\bottomrule
\end{tabular}
\end{table}

For the IM condition specifically, we identified two distinct approaches to using SI: corrective and additive (Table~\ref{tab:si-strategies}). Corrective SI (C) involves removing or modifying elements the user considers incorrect or irrelevant. Additive SI (A) involves adding new elements such as custom clusters or connections. We also note whether each strategy incorporated parametric exploration (P), as some participants combined SI with parameter-based refinement.

\begin{table}[!htb]
\caption{Semantic interaction strategies (C = Corrective SI, A = Additive SI, P = Parametric Exploration).}
\label{tab:si-strategies}
\centering
\small
\begin{tabular}{p{3cm}ccccp{8cm}}
\toprule
\textbf{Strategy} & \textbf{C} & \textbf{A} & \textbf{P} & \textbf{Type} & \textbf{Description} \\
\midrule
Removing Bias & $\times$ & & & Cumulative & Remove biased sources (e.g., clickbait) from the map \\
Topical Clusters & & $\times$ & & Single & Define custom topic clusters before exploration \\
Iterative Refinement & $\times$ & $\times$ & & Cumulative & Progressively refine through multiple interactions \\
Reset and Cluster & $\times$ & $\times$ & $\times$ & Mixed & Test clusters, reset, try different approach \\
Playground & $\times$ & $\times$ & $\times$ & Cumulative & Explore all interactions to test system limits \\
\bottomrule
\end{tabular}
\end{table}

A notable corrective strategy was ``Removing Bias,'' where participants systematically removed articles from sources they considered unreliable or sensationalistic. This shows how SI enables analysts to impose their own quality judgments on the extracted narrative. For additive strategies, ``Topical Clusters'' was particularly effective. Participants who defined custom clusters prior to exploration reported that it helped them organize their thinking and find more structured information. Qualitative feedback revealed that participants in both map conditions recognized the value of developing strategies over time, suggesting that extended use would yield additional benefits.

\subsection{Parameter Diversity}
Participants with the IM prototype showed less diversity in parameter settings compared to the BM prototype. Using the Gini coefficient as a diversity measure: for map size, BM = 0.25 vs. IM = 0.19; for coverage, BM = 0.29 vs. IM = 0.25; for temporal sensitivity, BM = 0.22 vs. IM = 0.19. Chi-square tests confirmed that the parameter distributions differed significantly between conditions (all $p < 0.001$).

Importantly, despite less direct parameter manipulation, IM participants generated a similar number of maps as BM participants (no significant differences in execution counts between map-based conditions). This pattern admits two competing explanations. First, SI may provide an alternative refinement pathway---a different mechanism for model refinement that reduces the need for parameter exploration while achieving similar exploration breadth. Second, the additional complexity of the SI capabilities may have increased cognitive load, reducing the participants' capacity for parameter exploration. The first interpretation is supported by the observation that IM users still generated similar numbers of maps despite less parameter diversity---if SI were not actively modifying the model, users would have had little reason to regenerate maps at comparable rates. However, fully distinguishing between these explanations would require direct measures of cognitive load, which we did not collect.

\subsection{Usability and Narrative Representation}
We evaluated perceived usability and narrative representation quality (see Table~\ref{tab:usability}). A multivariate analysis of variance (MANOVA) on the average Likert scores revealed a significant difference between groups ($p = 0.018$). Follow-up analysis showed that differences were concentrated in perceived Usefulness, Ease of Use, and Reuse.

\begin{table}[!htb]
\caption{Usability and narrative representation results (5-point Likert scale). (*) denotes items with significant differences.}
\label{tab:usability}
\centering
\small
\begin{tabular}{lccc}
\toprule
\textbf{Item} & \textbf{TL} & \textbf{BM} & \textbf{IM} \\
\midrule
Usefulness (*) & 4.10 & 4.31 & 4.80 \\
Ease of Use (*) & 4.50 & 4.00 & 3.60 \\
Reuse (*) & 4.10 & 4.38 & 4.90 \\
Representation & 4.10 & 3.92 & 4.20 \\
Redundancy & 2.90 & 2.85 & 3.60 \\
Relevance & 4.10 & 4.23 & 4.50 \\
Depth & 4.10 & 4.23 & 4.20 \\
\bottomrule
\end{tabular}
\end{table}

The IM prototype had the highest Usefulness and Reuse ratings, suggesting that despite requiring more effort to learn, the participants saw value in the SI capabilities for narrative sensemaking. The TL prototype had the highest Ease of Use but lowest Usefulness and Reuse scores. Map-based prototypes were perceived as more difficult to use than the timeline baseline, with IM showing the lowest Ease of Use rating---an expected trade-off given its additional interaction complexity. Importantly, there was no significant difference in narrative Representation ratings across conditions, indicating that all three approaches were perceived as equally capable of representing the underlying narrative structure.

\section{Discussion}
\label{sec:discussion}
\textbf{Narrative Maps and SI.} Our results provide evidence that map-based representations support sensemaking compared to timeline baseline, with the SI-enabled prototype showing the strongest performance. The IM condition significantly outperformed the timeline baseline in both high-level and detailed insights, while the BM showed intermediate performance. Qualitative findings support the value of SI, as participants developed distinct strategies (corrective and additive) that would be impossible without SI capabilities, and achieved comparable exploration breadth through different means than parameter manipulation. These findings are consistent with SI supporting \textit{incremental formalism}~\cite{shipman1999formalism}, although given the large observed effect sizes, even modest increases in sample size (approximately 20 per group) would likely suffice to detect these differences statistically. We also suggest longer tasks, which may potentially allow analysts to better familiarize themselves with the system and require more use of its advanced capabilities.

\textbf{Exploration Strategies and Usability Trade-offs.} The diversity of exploration strategies observed across map-based conditions (Table~\ref{tab:general-strategies}) suggests that narrative maps accommodate multiple sensemaking approaches. Some participants preferred a depth-first exploration of a single narrative structure, while others adopted breadth-first comparisons between alternative representations. This flexibility is valuable for narrative sensemaking, where analysts may have different goals or cognitive styles. However, this flexibility comes at a cost: the usability results reveal a trade-off between capability and complexity. The IM prototype scored highest on Usefulness and Reuse but lowest on Ease of Use, indicating that SI capabilities require additional learning effort.

This trade-off is consistent with the broader pattern observed in feature-rich VA tools, where increased analytical power often comes at the cost of initial learnability, which is a dynamic well-established in the Technology Acceptance Model~\cite{davis1989perceived}, which identifies perceived usefulness and perceived ease of use as competing factors in system adoption. Importantly, despite the lower Ease of Use rating, the IM prototype's highest Reuse scores suggest that participants perceived the long-term value of SI capabilities as outweighing the initial learning investment, indicating that the complexity barrier may diminish with extended use.

\textbf{Trust in SI.} Trust ratings were lower than usefulness ratings, although this difference was not statistically significant. We speculate that this pattern may relate to the ``With Respect to What'' problem in SI~\cite{wenskovitch2020respect}: user interactions can be ambiguous, and the system must choose how to interpret them across multiple underlying models. When these interpretations diverge from user intent, the resulting model updates may appear unpredictable, potentially undermining trust. This suggests that SI systems could benefit from explainability mechanisms that surface \emph{how} the system interpreted each interaction, allowing users to verify and correct misalignments. Confirming this hypothesis requires larger samples and direct measures of perceived interaction-interpretation alignment.

\textbf{Limitations.} Our participant pool consisted primarily of students, limiting generalizability to professional analysts. Unequal group sizes ($n = 10$ for TL and IM, $n = 13$ for BM) and a relatively small sample size limit the statistical power to detect differences between the IM and BM conditions. The duration of the task is shorter than that of real-world sensemaking tasks, and the 160-document dataset is smaller than many real-world collections. Given these constraints, significant IM-TL comparisons should be interpreted as evidence for map-based approaches in general; claims about SI-specific benefits are based primarily on qualitative evidence and require replication with larger samples.

Regarding scalability, the current narrative extraction pipeline involves a linear programming optimization whose complexity grows with the number of documents, and UMAP/HDBSCAN projections that may become less interpretable for very large collections. Previous work on the underlying narrative extraction framework has shown that it can handle datasets of several hundred documents~\cite{german2025trails}, and the interactive system has been designed with incremental model updates to support iterative refinement~\cite{keith2023iui}. Nevertheless, scaling to corpora of thousands of documents would likely require hierarchical extraction strategies~\cite{keith2026interactive} or sampling-based approaches. Future evaluations should test the system's performance and usability with larger, real-world document collections.

\section{Conclusions}
\label{sec:conclusions}
This paper presented a user study that evaluated narrative map representations and semantic interaction for sensemaking. Through a study with 33 participants, we demonstrated that map-based visualizations outperform timeline baselines for insight generation. While the SI-enabled prototype showed the strongest performance, the IM--BM differences did not reach statistical significance. However, the large observed effect sizes ($d > 0.8$) suggest that studies with approximately 20 participants per group would likely confirm SI-specific benefits.

Our qualitative findings identify distinct interaction strategies---corrective and additive---that analysts employ when SI is available, and reveal that SI provides an alternative to parameter exploration for model refinement. Additionally, the observed pattern of trust ratings suggests that the relationship between perceived system responsiveness and model trust merits further investigation, potentially through integration with explainability approaches.

Future work should investigate SI with larger pools of participants to better isolate its quantitative benefits, explore SI scalability to larger document collections, and develop tighter integration between SI and explainability to address interaction ambiguity.

\paragraph{Data and Code Availability.} The source code for the Narrative Maps Visualization Tool (NMVT) is publicly available at \url{https://github.com/briankeithn/narrative-maps}. The news dataset, survey instruments, and anonymized coded insights are available via Zenodo at: \url{https://doi.org/10.5281/zenodo.18930804}.

\begin{acknowledgments}
This research was supported by industry, government, and institute members of the NSF SHREC Center, which was founded in the IUCRC program of the National Science Foundation. This research was also supported by ANID/Doctorado Becas Chile/2019 - 72200105. Further analysis was supported by the ANID FONDECYT 11250039 Project "Interactive Narrative Analytics" and by Project 202311010033-VRIDT-UCN.
\end{acknowledgments}

\section*{Declaration on Generative AI}
During the preparation of this work, the authors used Grammarly and Writefull integrated with Overleaf to perform grammar and spelling corrections. Furthermore, Claude was used to improve writing and clarity in the camera-ready version. After using these tools/services, the authors reviewed and edited the content as needed and take full responsibility for the publication’s content. 

\bibliography{sample-ceur}


\end{document}